\documentclass[10pt,letterpaper]{article}
\usepackage{opex3}
\usepackage{color}
\usepackage{amsmath, amssymb, amsthm, wasysym}
\usepackage{graphicx, cite}
\usepackage{setspace}
\usepackage[symbol]{footmisc}

\newcommand{\icm}[1]{\ensuremath{#1\ \mathrm{cm^{-1}}}}
\newcommand{\fs}[1]{\ensuremath{#1\ \mathrm{fs}}}
\newcommand{\ps}[1]{\ensuremath{#1\ \mathrm{ps}}}
\newcommand{\nm}[1]{\ensuremath{#1\ \mathrm{nm}}}

\graphicspath{{Figures/}}

\begin{document}

\title{Optimal all-optical switching of a microcavity resonance in the telecom range using the electronic Kerr effect}

\author{Emre Y\"uce,$^{1,*}$ Georgios Ctistis,$^{1}$ Julien Claudon,$^{2,3}$ \\ Jean-Michel G\'erard,$^{2,3}$ and {Willem L. Vos}$^{1}$}

\address{
$^{1}$Complex Photonic Systems (COPS), MESA+ Institute for Nanotechnology, University of Twente, P.O. Box 217, 7500 AE Enschede, The Netherlands \\
$^{2}$Univ. Grenoble Alpes, INAC-SP2M, ``Nanophysique et Semiconducteurs" group, 38000 Grenoble, France\\
$^{3}$CEA, INAC-SP2M, ``Nanophysique et Semiconducteurs" group, 38000 Grenoble, France}

\email{$^{*}$e.yuce@utwente.nl} 



\begin{abstract}
We have switched GaAs/AlAs and AlGaAs/AlAs planar microcavities that operate in the ``Original" (O) telecom band by exploiting the instantaneous electronic Kerr effect. 
We observe that the resonance frequency reversibly shifts within one picosecond. 
We investigate experimentally and theoretically the role of several main parameters: the material backbone and its electronic bandgap, the pump power, the quality factor, and the duration of the switch pulse.
The magnitude of the shift is reduced when the backbone of the central $\lambda-$layer has a greater electronic bandgap; pumping with photon energies near the bandgap resonantly enhances the switched magnitude. 
Our model shows that the magnitude of the resonance frequency shift depends on the pump pulse duration and is maximized when the duration matches the cavity storage time that is set by the quality factor. 
We provide the settings for the essential parameters so that the frequency shift of the cavity resonance can be increased to one linewidth.
\end{abstract}


\bibliographystyle{osajnl}

\section{Introduction}
Fast switching rates are currently under demand by both optical information technologies~\cite{kanan.1996.ieee, mondia.2005.josab, altug.2006.nat.phys, fushman.2007.apl, harding.2007.apl, husko.2009.apl, wada.2014.jpd, radic.2014.oe, yang.2014.pra} and by fundamental studies that aim to manipulate light-matter interactions at femtosecond time scales~\cite{johnson.2002.prb, yanik.2004.prl, henri.2013.oe, fiore.2015.pra}. 
The electronic Kerr effect inherently provides the highest possible speed given its virtually instantaneous response nature \cite{boyd.2008.book,hartsuiker.2008.jap, harding.2009.josab, ctistis.2011.apl, yuce.2012.josab, rossi.2012.oe, yuce.2013.ol, wada.2014.jpd}. 
Using the Kerr effect the resonance of a microcavity has been switched within a duration as short as \fs{300}~\cite{ctistis.2011.apl}, and repeated switching has been performed at unprecedented single-channel rates beyond one THz clock rate~\cite{yuce.2013.ol}. 
The electronic Kerr effect is a third-order nonlinear process and its magnitude increases linearly with the intensity of the pump laser pulse~\cite{boyd.2008.book}. 
Increasing the intensity of the laser pulse, however, causes the excitation of free carriers that have a much slower speed and counteract the electronic Kerr effect~\cite{yuce.2012.josab, li.2013.ieee}. 
Fortunately, the excitation of free carriers can be suppressed via the judicious selection of the photon energy and the intensity of the switching pulse~\cite{hartsuiker.2008.jap, harding.2009.josab, ctistis.2011.apl, yuce.2012.josab, yuce.2013.ol}. 

In switching the cavity resonance by the electronic Kerr effect a main challenge remains, namely how to increase the shift of the cavity resonance beyond one linewidth. 
To overcome this challenge one has to delicately choose all parameters that play a role in frequency shift of the cavity resonance that involves; 1) the backbone and the frequency of light relative to the backbone's electronic bandgap, 2) the intensity of the light pulses, 3) the quality factor of the cavity, and 4) the duration of the switch pulse. In this work, we explore these crucial parameters and provide a method to maximize the resonance frequency shift induced by the electronic Kerr effect.

\section{Experimental details}

\subsection{Samples}

We have performed experiments on planar microcavities that consist of a GaAs $\lambda$-layer ($d=376 \ \rm{nm}$) sandwiched between two Bragg stacks consisting of 15 and 19 pairs of $\lambda/4$-thick layers of nominally pure GaAs ($d_{GaAs}=94 \ \rm{nm}$) and AlAs ($d_{AlAs}= 110 \ \rm{nm}$), respectively, that are grown on a GaAs wafer. 
Figure \ref{yuce-f1}(b) shows a scanning electron micrograph (SEM) cross-section of a GaAs/AlAs sample. 
Since the bottom Bragg mirror is positioned on a GaAs wafer there is a smaller refractive index contrast that results in a lower reflectivity. 
Therefore, a greater number of layers is required for the bottom Bragg stack to achieve a similar reflectivity as for the upper Bragg stack. 
The cavity resonance is designed to occur at $\lambda_0=1280 \pm 5 \ \rm{nm}$ in the Original (\emph{O}) telecom band. 

\begin{table}[htb]
\begin{center}
\caption{List of samples used in this work. The resonance frequency $\omega_0$ and corresponding wavelength $\lambda_0$, and the quality factor $Q$ of the cavities are obtained from our measurements. The last column shows in which sections the cavities are discussed.}  
\begin{tabular}{l l l l l l rp{4cm}}  
\hline\hline \\ [0.1ex]                     
Quality & Backbone & Top/Bottom & $\omega_{0}$ [$\mathrm{cm^{-1}}$] & $\lambda_{0}$ $[\mathrm{nm}]$ &Used in  \\ 
 factor & & Num. of pairs & & & Section \\[1.0ex]   
\hline \hline \\ [0.1ex]              
$390 \pm 60$ & GaAs/AlAs &7/19 &$7806 \pm 40$ & $1281 \pm 6$& 3.1, 3.2, 3.3 \\
$540 \pm 60$ & GaAs/AlAs &11/19 &$7762 \pm 40$ &  $1288 \pm 6$& 3.3 \\

$890 \pm 60$ & GaAs/AlAs &15/19 &$7806 \pm 40$ &   $1281 \pm 6$& 3.3 \\

$210 \pm 60$ & $\rm{Al_{0.3}Ga_{0.7}As}$/AlAs & 9/16&$8038 \pm 40$ &  $1244 \pm 6$& 3.1 \\
\hline \hline
\end{tabular}%
\label{table1}
\end{center}
\end{table}

In order to achieve a range of quality factors we prepared a large sample and cut it into smaller chips ($ 5~\rm{mm} \times 5~\rm{mm}$). 
Next, a number of layers is selectively removed from the top Bragg stack of one chosen chip by dry and wet etching techniques to obtain four GaAs/AlAs cavities with sequentially reduced quality factors. 
Table~\ref{table1} lists all samples that are studied in this work. 

Figure~\ref{yuce-f1}(c) shows the measured and the calculated reflectivity spectra of the GaAs/AlAs microcavity. 
The reflectivity spectra of the cavity is measured with a setup consisting of a supercontinuum white-light source (Fianium) and a Fourier-transform interferometer with a resolution of $0.5 \ \rm{cm^{-1}}$ (BioRad FTS6000). 
We see that the stopband of the Bragg stacks extends from approximately $7050 \ \rm{cm^{-1}}$ to $8500 \ \rm{cm^{-1}}$ ($1418.4 \ \rm{nm}$ to $1176.5 \ \rm{nm}$). 
On both sides of the stopband Fabry-P\'erot fringes are visible due to interference of the light that is reflected from the front and the back surfaces of the sample. 
Inside the stopband a narrow trough corresponds to the cavity resonance. 

To investigate the effect of the bandgap of the backbone on Kerr switching we have also studied a planar microcavity made of a $\rm{Al_{0.3}Ga_{0.7}As}$ $\lambda$-layer ($d=400 \ \rm{nm}$) sandwiched between two Bragg stacks made of 9 and 16 pairs of $\lambda/4$-thick layers of $\rm{Al_{0.3}Ga_{0.7}As}$ ($d_{AlGaAs}=100.2 \ \rm{nm}$) and AlAs ($d_{AlAs}= 111.7 \ \rm{nm}$), respectively, and grown on a GaAs wafer. 
The AlGaAs/AlAs cavity is designed to resonate at $\lambda_0=1280 \pm 5 \ \rm{nm}$ in the Original (\emph{O}) telecom band and has a quality factor factor of $Q=210$.

\subsection{Setup} \label{setup}

\begin{figure}[htb]
\begin{center}
\includegraphics[width=0.75\columnwidth]{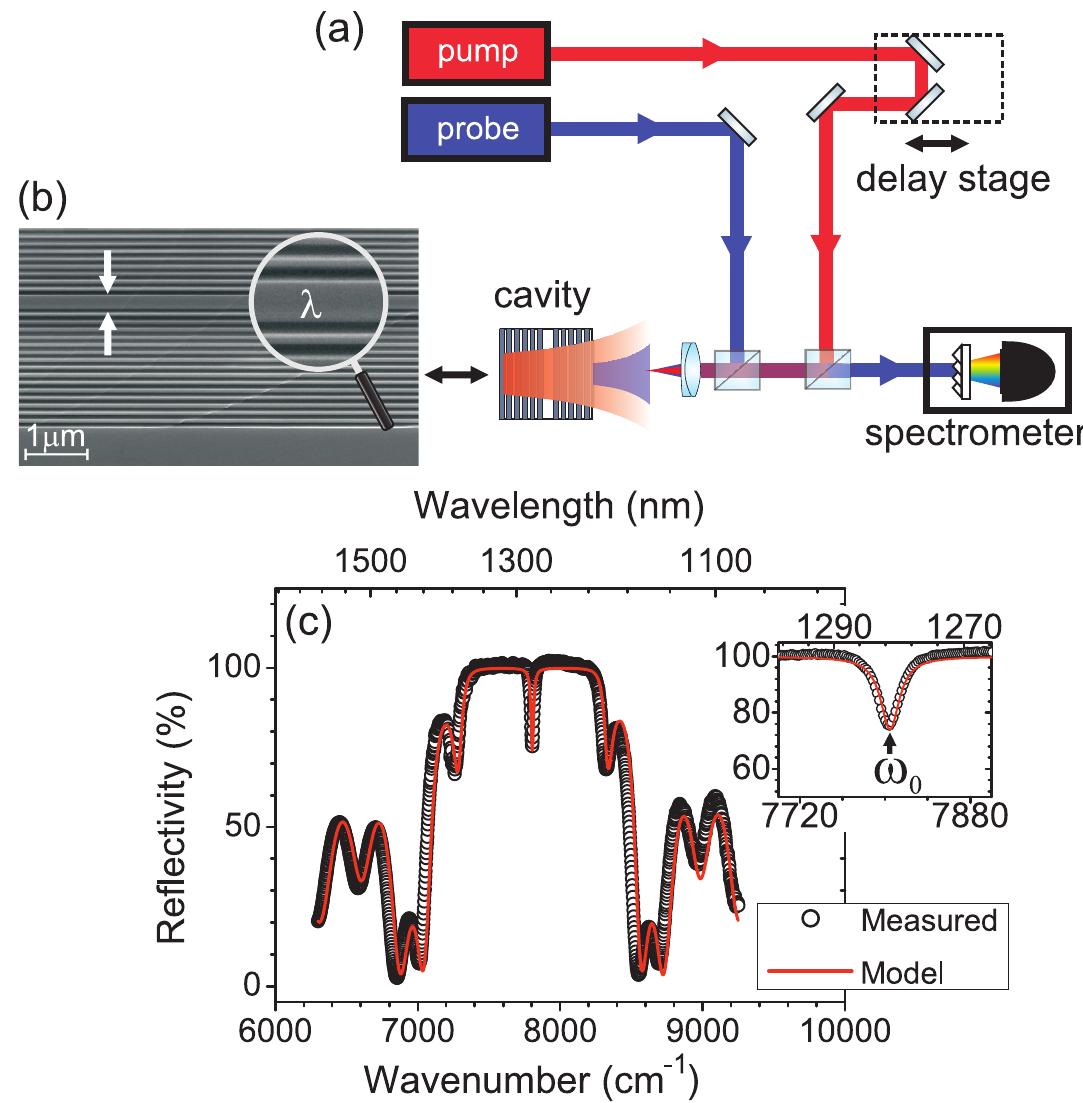}
\caption{\emph{(color)  
(a) Schematic of the switch setup. 
The probe beam path is shown in blue, the pump beam path in red. 
The time delay between pump and probe pulses is set with a delay stage. 
The reflected signal from the cavity is spectrally resolved and detected.
(b) SEM picture of the multilayer structure of a GaAs/AlAs microcavity. 
GaAs layers appear light grey, and AlAs layers dark grey. 
The white arrows indicate the thickness of the GaAs $\lambda$-layer. 
The GaAs substrate is seen at the bottom. 
The magnifier shows how the $\lambda$-layer is sandwiched between the Bragg stacks. 
(c) Measured (black symbols) and calculated (red line) reflectivity spectra of a GaAs/AlAs microcavity with $Q=390 \pm 60$. The vertical axes show the reflectivity, the horizontal axes show the frequency (bottom) and the wavelength (top). Within the stopband a narrow trough indicates the cavity resonance frequency $\omega_0$, shown with higher resolution in the inset. The calculations are performed with a transfer matrix model.
}}
\label{yuce-f1}
\end{center}
\end{figure}

A versatile ultrafast pump-probe setup is used to Kerr-switch our microcavity~\cite{euser.2009.rsi}. 
The setup is shown in Fig. \ref{yuce-f1}(a) and consists of two independently tunable optical parametric amplifiers (OPA, Light Conversion Topas) pumped by a 1 kHz oscillator (Hurricane, Spectra Physics) that are the sources of the pump and probe beams. 
The pulse duration of both OPAs is $\tau_{P} = 140 \pm 10 \ \rm{fs}$.  
The time delay $\Delta t$ between the pump and the probe pulse is set by a delay stage with a resolution of $15 \ \rm{fs}$.  
The measured transient reflectivity contains information on the cavity resonance during the cavity storage time and it should thus not be confused with the instantaneous reflectivity at the delay $\Delta t$. 
We explain the transient reflectivity using the description given in Refs.~\cite{euser.2009.rsi, harding.thesis, hartsuiker.thesis}.
In the absence of spectral filtering, the measured signal J, neglecting electronic amplification factors, is equal to the magnitude of the time- and space integrated Poynting vector S~\cite{euser.2009.rsi, harding.thesis}:
\begin{eqnarray}
J=\pi r^2 \int_{-t_{int}/2}^{t_{int}/2} |S| dt &=& \int_{-t_{int}/2}^{t_{int}/2} \sqrt{\frac{\epsilon_0}{\mu_0}}E(t)^2dt 
\label{step1} \\ 
&\approx & \pi r^2 \sqrt{\frac{\epsilon_0}{\mu_0}} \frac{\tilde{E_0}}{2} \int_{-\infty}^{\infty}(e^{-4 ln 2 t^2/\tau^2_{P} })^2 dt \label{step2} \\ 
&=& \pi r^2 \sqrt{\frac{\epsilon_0}{\mu_0}} \sqrt{\frac{\pi}{2ln(2)}} \frac{\tau_{P} \tilde{E_0}^2}{4},
\label{TransientSignal}
\end{eqnarray}
where the electric field E(t) reflected by a mirror onto the detector can be separated in a Gaussian envelope $\tilde{E}(t)$ of FWHM $ \tau_{P} $ and amplitude $\tilde{E}_0$ that is multiplied by sinusoidal component with a carrier frequency $\omega_0$ in rad/s~\footnote{This slowly varying envelope Approximation (SVEA, see e.g.~\cite{Rudolph.1996}) can be applied to pulses where $\tau_{P} >> 1/\omega_0$, and where $\omega_0$ does not change over t, i.e., for bandwidth limited pulses. 
For pulses whose envelope is broadened by the interaction with a cavity, the analytic expression (Eq.~\ref{TransientSignal}) is not valid, but the approximation of the integration limits remains the same.}. 
The beam is collimated and has radius $r$, 
$\epsilon_0$ and $\mu_0$ denote the permitivity and permeability of free space, respectively. 
Since the integration time $t_{int}$ of the InGaAs line array detector is much longer than any probe interaction time $\tau_{pr}$~\footnote{The probe interaction time is either $\tau_{pr}=\tau_{P}$ or $\tau_{pr}=Q/\omega_0$, whichever is greater, and it is in the 100 fs to 1 ps range.}, we essentially integrate all probe light that is stored or reflected by the cavity, given a pump-probe time delay $\Delta t$. 
Therefore, the boundaries of time integral in Eq.~\ref{step1} can be taken to be infinity because $t_{int} >> \tau_P$. 
The squared oscillating term can then be integrated separately and yields $1/2$. 
In Eq.~\ref{step2} we approximate the peak intensity for a focussed Gaussian pulse as $I=4\sqrt{ln2}G / (\pi^{3/2} r^2 \tau_{P})$, where r is the waist radius at the focus and G the energy per pulse.
Eq.~\ref{step2} reveals that it is not the instantaneous transmission or reflection that is measured, but the integrated intensity. 

In our study, we use a spectrometer to frequency resolve the reflected transient signal. 
The reflected signal from the cavity is spectrally filtered with a spectrometer (Acton) and detected with a nitrogen cooled InGaAs line array detector (Princeton Instruments).
Therefore, the observed spectrum, without amplification and conversion factors, is a Fourier transform of $E(t)$\cite{hartsuiker.thesis, harding.thesis} 
\begin{equation}
J(\omega)=\pi r^2 (\epsilon_0 c)^{-1} \left| \int_{-\infty}^{\infty} dt E(t) e^{i\omega t} \right|^2
\end{equation}
where c is the velocity of light in free space. 
The field escaping from a cavity whose resonance frequency shifts in time may exhibit new frequency components where the amplitude is higher than that of the incident bandwidth limited pulse. 
In this case, the ratio of the reflected pulse to the reference pulse, called the transient reflectivity $R^t(\omega)=J(\omega)_{sample}/J(\omega)_{ref}$  exceeds unity at the new frequency components. 
In this sense, the transient reflectivity differs from the reflectivity measured in a CW experiment that is necessarily always bounded to $100 \%$.
The measured transient reflectivity $R^t(\omega)$ is a result of the probe light that impinges at delay $\Delta t$, circulates in the cavity during on average the storage time $\tau_c$, escapes, and is then integrated by the detector. 
Therefore, we call the measured signal the transient reflectivity or the transient transmission.

The cavity is switched with the electronic Kerr effect by judicious tuning of the pump and the probe frequencies relative to the semiconductor bandgap of the cavity backbone~\cite{hartsuiker.2008.jap, harding.2009.josab}.  
The probe frequency $\omega_{pr}$ is set by the cavity resonance in the telecom range (see Table~\ref{table1}).   
Furthermore, the photon energy of the pump light is chosen to lie below half of the semiconductor bandgap energy of GaAs and AlGaAs($E_{pu}<\frac{1}{2}E_{gap}$), see Figs. \ref{yuce-f4}(c) and (d), to avoid two pump-photon excitation of free carriers. 
Therefore, the pump frequency is centered at $\omega_{pu}= 4165 \ \rm{cm^{-1}}$ $(\lambda_{pu}=2400 \rm{nm})$. 
The frequency of the pump and the probe light is kept the same for the GaAs and  AlGaAs cavities to directly compare the effect of photon energy relative to the electronic bandgap. 
The probe fluence is set to $I_{pr}=0.18 \pm0.02 \rm{\ pJ / \mu m^2}$ while the average pump fluence is set to $I_{pu}=65 \pm 20 \rm{\ pJ / \mu m^2}$. 
The fluences are chosen as they yield an as large as possible Kerr effect without unwanted free carrier excitation~\cite{yuce.2012.josab}.
The fluence of the pulses is determined from the average laser power at the sample position and converted to peak power assuming a Gaussian pulse shape. The pump beam has a larger Gaussian focus ($2r_{pu}= 70 \ \rm{\mu m}$) than the probe beam ($2r_{pr}=30 \ \rm{\mu m}$) to ensure that only the central flat part of the pump focus is probed and that the probed region is spatially homogeneously pumped~\cite{euser.2005.jap}. 
The induced resonance frequency shift with the electronic Kerr effect is determined by the temporal($\Delta t$) and the spatial overlap of the pump and the probe beams. 
For this reason, once we fix the spatial alignment of the pump and probe beams we successively perform switching of the different cavities to allow for the best possible comparison.

\section{Model to calculate time-resolved spectra} \label{Model}

The model that we employ to calculate the time-resolved transient reflectivity $R^t$ spectra has previously been described briefly in Ref.~\cite{harding.2012.josab}. 
The probe field is calculated in the time domain at every position in a one-dimensional planar microcavity that experiences a time-dependent refractive index. 
To account for the induced refractive index change $n(t)$, we consider here the positive non-degenerate Kerr coefficient of GaAs~\cite{yuce.2012.josab}. 
We start with a Gaussian probe pulse at $z = z_0$:
\begin{equation}
E_{pr}(z_0,t)=E_0(z_0) e^{-i\omega t} e^{-({t-t_0}/\tau_{pr})^2},
\label{probebefore}
\end{equation}
where $\mathrm{E_0}$ is the amplitude and $\omega$ is the angular frequency of the probe. 
In our calculations we chose a short duration for the probe pulse ($\tau_{pr}=\fs{10}$) to obtain a broad spectral bandwidth and thus a flat response within the spectral region of the cavity resonance. 
The field that starts from $z = z_0$ travels in homogeneous medium with a time dependent refractive index $n(t)$. 
The time that it takes for the field to travel from position $z_0$ to $z$ is then equal to $t=n(t)\cdot(z-z_0)/c$. 
As a result, the Gaussian pulse at position $z$ is given by 
\begin{eqnarray}
E_{pr}(z,t)&&=E_{pr}(z_0,n(t)\cdot(z-z_0)/c)\nonumber \\
&&=E_0(z_0) e^{i\omega(n(t)\cdot(z-z_0)/c)} e^{-(({n(t)\cdot(z-z_0)/c)-t_0}/\tau_{pr})^2}.
\label{fieldat}
\end{eqnarray}

\begin{figure}[htb]
\begin{center}
\includegraphics[width=0.5\columnwidth]{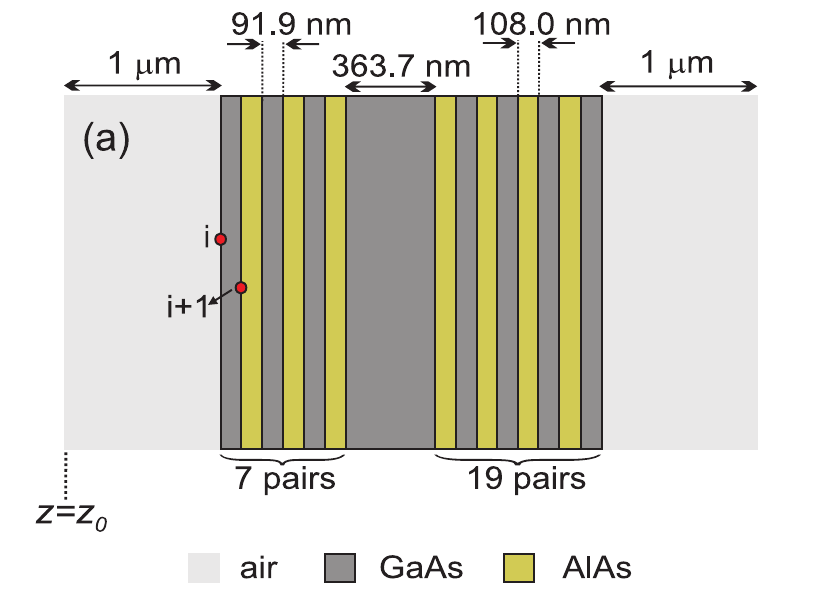}
\caption{Schematic picture of the one-dimensional microcavity considered in our model calculations. 
The Bragg mirrors consist of GaAs and AlAs layers and the $\lambda$-layer consists of GaAs. 
The thickness of the air, GaAs, and AlAs layers are indicated in the figure so as to yield a resonance frequency as in our experiment. 
The first two interfaces are marked with the indices i and i+1. 
The probe field is launched at $z = z_0$. }
\label{yuce-f2}
\end{center}
\end{figure}

Similar to our experiments the structure that we describe in our model consists of air, the top Bragg mirror, the $\lambda$-layer, the bottom Bragg mirror, and air after the cavity structure, as shown in Fig.~\ref{yuce-f2}. 
Since the thickness of the GaAs wafer is not exactly known, we exclude it in our model.
The Bragg mirrors consist of AlAs and GaAs layers with unswitched refractive indices $\rm{n_{AlAs} \ and \ n^0_{GaAs}}$, respectively. 
During the switching of our microcavity we take the refractive index of AlAs and air to be constant in time, whereas the refractive index of GaAs is time dependent. 
The refractive index of GaAs changes with the time delay $\Delta t$ between the pump and the probe pulses due to the electronic Kerr effect. 
Hence, we define a position and time dependent refractive index for the structure as follows: 
\begin{eqnarray}
n(z,t) = \begin{cases} n_{air} &  \mbox{, z in air} 
\\ n_{AlAs} &  \mbox{, z in AlAs}
\\n^0_{GaAs}+ \frac{12 \pi^2 \chi^{(3)}}{(n^0_{GaAs})^2 c}  
\cdot [ I_{pu}e^{-(\frac{t-\Delta t}{\tau_{pu}})^2}] &  \mbox{, z in GaAs}, \\
\end{cases}
\label{RefIndxChnge}
\end{eqnarray}
where $\chi^{(3)}$ is the third-order susceptibility of GaAs, $I_{pu}$ the peak intensity of the pump pulse, and $\tau_{pu}$ the duration of the pump pulse. 
In our calculations we neglect the refractive index change induced by the probe light since the intensity of the probe is much lower than the pump intensity.
Since the probe field propagates through the microcavity structure that consists of many different materials with different refractive indices, the field encounters many interfaces. 
The field that impinges on interface is partly reflected and partly transmitted, as given by the Fresnel coefficients~\cite{born.2002}. 
The reflected ($\mathfrak{r}$) and transmitted ($\mathfrak{t}$) amplitude coefficients at normal incidence from any interface are equal to 
\begin{eqnarray}
\mathfrak{r}=\frac{n_1(z,t)-n_2(z,t)}{n_1(z,t)+n_2(z,t)}, \nonumber \\
\mathfrak{t}=\frac{2 n_1(z,t)}{n_1(z,t)+n_2(z,t)},
\label{fresnel}
\end{eqnarray}
where $n_1(z,t)$ and $n_2(z,t)$ are the time-dependent refractive indices of the first and the second medium, respectively. 
Due to the transmission and reflection from an interface, there are fields travelling in opposite directions. 
Part of the field transmitted by interface $i$ is reflected from the next interface $i+1$ and thus interferes with the incident field. 
As a result, at a given position $z$ inside the microcavity, the field is equal to 
\begin{eqnarray}
{\cal E}^{i}_{pr}(z,t)&&=E^+_{pr}(z_0,n(z,t)\cdot(z-z_0)/c) \cdot \mathfrak{t}^{i} \nonumber \\ 
&&+ E^-_{pr}(z_0,n(z,t)\cdot(z-z_0)/c) \cdot \mathfrak{r}^{i+1}
\label{fresnelfieldat}
\end{eqnarray}
For convenience we take the direction of the transmission as the positive direction. 
Since the microcavity structure consists of $N$ interfaces, we have generalized Equation~\ref{fresnelfieldat} to the N relevant interfaces. 
 
We calculate the field at any position $z$ in the multilayer structure by inserting the time-dependent refractive index of GaAs and the time-independent refractive indices of AlAs and air in $n(z,t)$ from Eq.~\ref{RefIndxChnge} into Eq.~\ref{fresnelfieldat}. 
To calculate the transient reflectivity spectrum we include all interfaces, see Fig.~\ref{yuce-f2}, to obtain the total time-resolved field ${\cal E}_{pr}(z,t)$. 
Next, we perform a discrete Fourier transform on such a field in reflection geometry 
\begin{equation}
\vert {\cal E}_{pr}(z, \omega) \vert^2=\bigg\vert \sum^{t}_{0}{\cal E}_{pr}(z, t)\cdot e^{-(i2\pi \omega \delta t)} \bigg\vert^2. 
\label{fourier}
\end{equation}
to obtain the transient field ${\cal E}(z, \omega)$, and thereby the transient reflectivity $R^t$ spectra. 

\begin{figure}[htb]
  \centering
  \includegraphics[width=0.95\columnwidth]{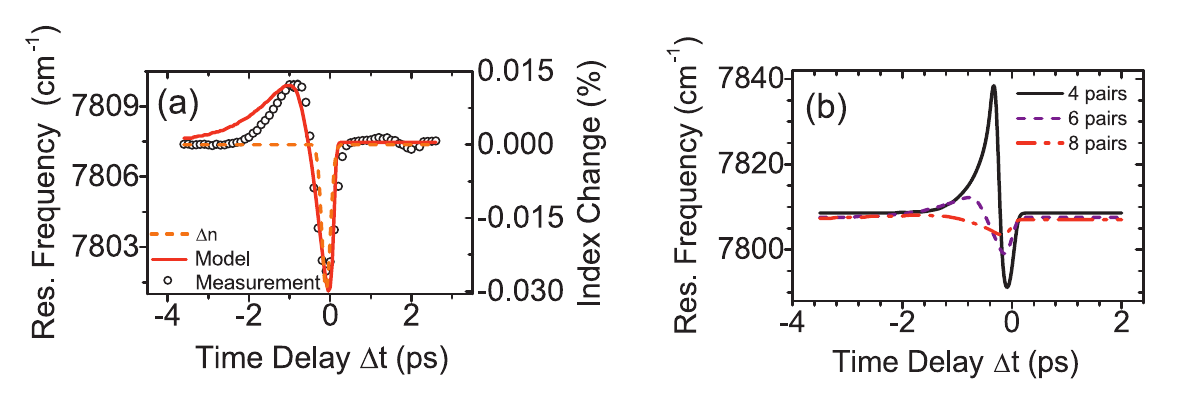}
  \caption{(a) Measured (symbols) and calculated (red curve) resonance frequency versus time delay ($\Delta t$) between pump and probe for a GaAs/AlAs cavity ($Q=390 \pm 60$). 
The resonance frequency red-shifts due to increased refractive index only near temporal overlap ($\Delta t=0 \ \pm 15 \ fs$) of pump-probe, shown with dashed curve. 
(b) Calculated spectra for a GaAs/AlAs cavity that consists of 19 pairs of bottom Bragg layers, $\lambda-$layer, and  sequentially changed number of layers on the top Bragg mirror.
}
\label{yuce-f3}
\end{figure}

We model our experimental results using our dynamic model, presented in Fig.~\ref{yuce-f3}. 
Figure~\ref{yuce-f3}(a) shows the resonance frequency versus pump-probe time delay $\Delta t$ for the GaAs/AlAs cavity with $Q=390 \pm 60$. 
The resonance is taken as the minimum of the transient reflectivity trough, see Fig.~\ref{yuce-f1}(c).
The resonance quickly shifts by $5.6 \ \rm{cm^{-1}}$ to a lower frequency at pump-probe overlap ($\Delta t=0$) and quickly returns to the starting frequency within \ps{1}. 
The shift of the cavity resonance to a lower frequency is due to the increased refractive index of GaAs, shown on the right ordinate. 
Our dynamic model predicts the frequency shift during the instantaneous switching of the cavity in excellent agreement with our experimental results.
Figure~\ref{yuce-f3}(a) shows that the minimum of the resonance trough appears at a higher frequency when the probe pulse arrives before the pump pulse ($\Delta t<-500$ fs) even though the refractive index only increases. 
The apparent blue shift appears since we track the minimum of the cavity trough (see inset of Fig.~\ref{yuce-f1}), which is not necessarily the instantaneous resonance frequency.
The instantaneous resonance frequency, which shifts within the duration of the pump pulse $\tau_P$, follows the refractive index change that is also shown in Fig.~\ref{yuce-f3}(a).
The apparent blue shift of the cavity trough is a result of the asymmetric cavity design. 
In the asymmetric cavities the top Bragg mirror consist of fewer layers than the bottom Bragg mirror. 
This results in more leakage from the top Bragg mirror. 
As a result, the interference between the probe light that escapes from the cavity (where it is influenced by the pump pulse) and the probe light that has directly reflected from the top mirror increases. 
Given the increased modulation, the resonance trough seems to appear at a higher frequency, even though the refractive index does not yet change. 
In order to explain the effect of the asymmetry we have performed calculations for cavities with sequentially increased number of top Bragg layers as shown in Fig.~\ref{yuce-f3}(b).
With increasing number of top layers the cavity becomes more symmetric and at the same time the quality factor of the cavity increases.
In Fig.~\ref{yuce-f3}(b) we observe that the apparent blue shift of the cavity resonance at $\Delta t<0$ decreases for increasingly symmetric cavities.
In Fig.~\ref{yuce-f3}(b) the red shift of the cavity resonance at $\Delta t=0$ decreases with increasing quality factor, which will be explained in section~\ref{SecQfactor}.

\section{Maximizing the frequency shift induced via the electronic Kerr effect}

\subsection{The effect of the backbone's electronic bandgap}

\begin{figure}[htb]
  \centering
  \includegraphics[width=0.9\columnwidth]{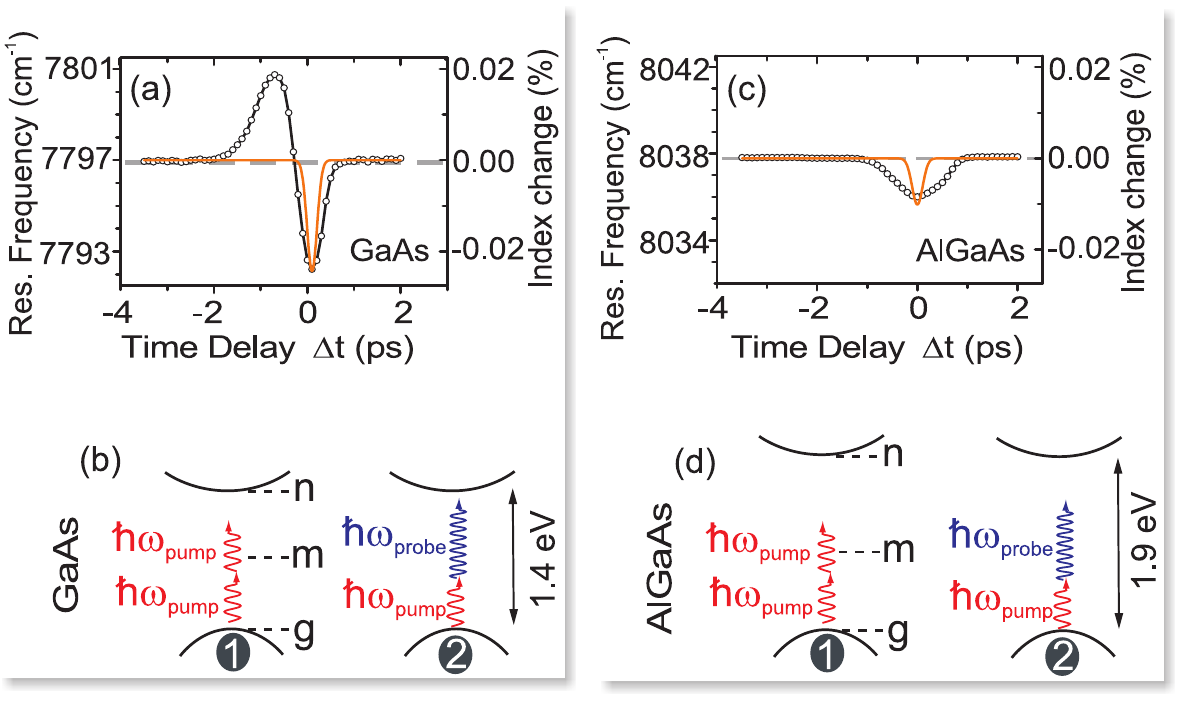}
  \caption{Resonance frequency versus time delay ($\Delta t$) between pump and probe for (a) GaAs/AlAs and (c) AlGaAs/AlAs cavity. The resonance frequency red-shifts due to increased refractive index only near temporal overlap ($\Delta t=0 \ \pm 15 \ fs$) of pump-probe. Both cavities are switched at $65 \ \rm{pJ/\mu m^2}$ pump fluence. The dashed lines represent the unswitched cavity resonance frequency. The solid curves represent the induced refractive index change. The schematic representation of the electronic bandgap of (b) GaAs and (d) AlGaAs and the energy of the pump and probe photons relative to the bandgap.}
\label{yuce-f4}
\end{figure}

To compare the effect of the backbone's electronic bandgap we have performed Kerr switching experiments on cavities that consist of GaAs/AlAs and AlGaAs/AlAs. 
Figure~\ref{yuce-f4} shows the resonance frequency versus time delay for the GaAs/Alas and AlGaAs/AlAs cavities. 
The cavity resonance for the AlGaAs cavity shifts by \icm{1.8}, which is less than \icm{4.7} of the GaAs cavity. 
The observation of a smaller frequency shift for the AlGaAs backbone is surprising in view of the fact that the nonlinear refractive index coefficient $n_2^{AlGaAs}=1.55 \times 10^{-4} \ \mathrm{cm^2/GW}$~\cite{caspani.2011.josab} is greater than of GaAs  $n_2^{GaAs}=0.5 \times 10^{-4} \ \mathrm{cm^2/GW}$~\cite{hartsuiker.2008.jap}. 
To understand this lower frequency shift with AlGaAs/AlAs cavity we consider how the third order susceptibility depends on material parameters. 

\begin{figure}[htb]
  \centering
  \includegraphics[width=0.6\columnwidth]{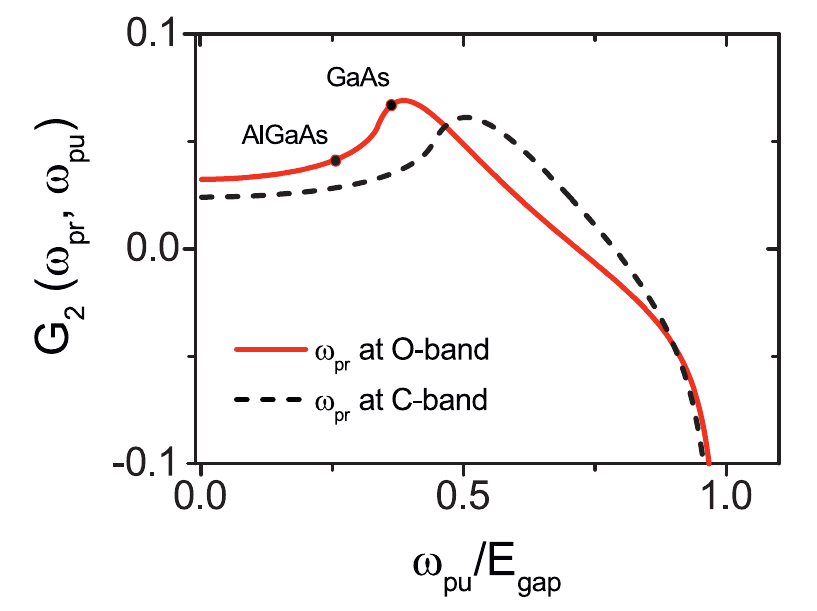}
  \caption{Nondegenerate dispersion curve of the electronic Kerr effect for probe frequency within original $(O)$ and conventional $(C)$ telecom bands, shown with solid and dashed curves, respectively. The symbols mark the $G_2$ values at our setting of pump frequency for GaAs and AlGaAs cavities.}
\label{yuce-f5}
\end{figure}

Figure~\ref{yuce-f5} shows the nondegenerate dispersion curve of the electronic Kerr effect for probe frequency $\omega_{pr}$ within the original $(O)$ and conventional $(C)$ telecom bands. 
The dispersion curve is calculated using the method described by Sheik-Bahae et al.~\cite{sheik-bahae.1991.ieee}. 
The dispersion of the nonlinear refractive index coefficient $n_2$ has been validated experimentally in Refs.~\cite{kuzyk.1989.josab, sheik-bahae.1990.prl, sheik-bahae.1991.ieee, hurlbut.2007.ol}. 
In Fig.~\ref{yuce-f5} we see that the nonlinear index coefficient is maximized near the nondegenerate two photon absorption edge~\cite{sheik-bahae.1990.prl, sheik-bahae.1991.ieee}. 
Our cavities are designed to operate within the original $(O)$ telecom band, which corresponds to a reduced probe frequency $\omega_{pr}/E_{gap}=0.65$.
We set the pump frequency to $\omega_{pu}/E_{gap}=0.35$ in order to suppress degenerate free carrier excitation, see Fig.~\ref{yuce-f4}(b1).
The non-degenerate free carrier excitation (pump and probe, Fig.~\ref{yuce-f4}(b2)) is suppressed since the probe fluence is much smaller than the pump fluence. 
At this setting of the pump frequency, the non-degenerate sum of pump and of probe frequency are tuned close to the electronic bandgap of the material.
As a result, the nonlinear index coefficient is close to the maximum, as shown in Fig.~\ref{yuce-f5}. 
We use the same frequency of the pump and of the probe light for the AlGaAs/AlAs cavity. 
In this case we operate away from the electronic bandgap of AlGaAs both for degenerate two-pump photon excitation (Fig.~\ref{yuce-f4}(d1)) and for non-degenerate pump and probe photon excitation (Fig.~\ref{yuce-f4}(d2)). 
Consequently, we observe less refractive index change due to a smaller nonlinear refractive index coefficient, see Fig.~\ref{yuce-f5}, which explains the smaller shift of the cavity resonance in Fig.~\ref{yuce-f4}(c).

Figure~\ref{yuce-f5} also shows the dependence of the nonlinear index coefficient versus pump frequency when the cavity resonance is set to operate within $C$-band.
In this case, we see that the electronic Kerr effect is maximized when the pump frequency is tuned to $\omega_{pu}/E_{gap}=0.5$ ($\lambda_{pu} \simeq \nm{1700}$ for GaAs).
At this pump frequency, however, the excitation of free carriers via two pump photons will be strong, which will hinder the electronic Kerr effect~\cite{yuce.2012.josab}.
Thus, we conclude that cavities operating within $O$-band are more amenable for ultrafast switching using the electronic Kerr effect, since the nonlinear index coefficient can be maximized while suppressing free carrier excitation.

\subsection{The effect of the pump power}

\begin{figure}[htb]
  \centering
  \includegraphics[width=0.6\columnwidth]{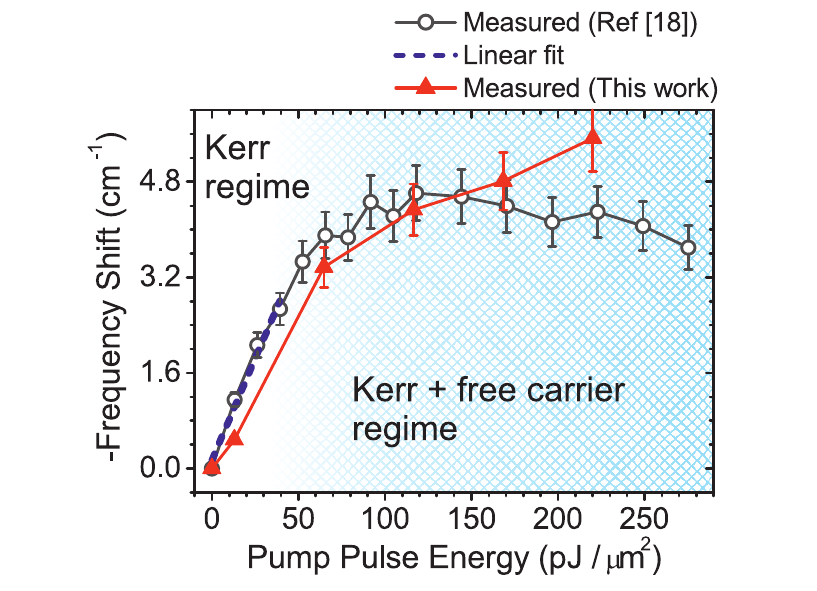}
  \caption{Instantaneous shift of the resonance frequency versus pump-pulse energy for a GaAs/AlAs cavity ($Q=390 \pm 60$) at pump and probe frequencies $\omega_{pu} = \icm{4165}$ and $\omega_{pr} = \icm{7806}$, respectively. Red triangles show the measured results and the line is a guide to the eye. At low pump-pulse energies we only observe electronic Kerr effect. The competing free carrier regime is extended beyond $70 \ pJ / \mu m^2$ and the region with both Kerr and free carrier excitation is shaded. The black symbols show the data from Ref.\cite{yuce.2012.josab}.}
\label{yuce-f6}
\end{figure}

To increase the resonance frequency shift induced by the electronic Kerr effect one could naively increase the pump pulse energy. 
Figure~\ref{yuce-f6} shows the shift of the cavity resonance frequency versus pump-pulse energy for GaAs/AlAs cavity with $Q=390 \pm 60$. 
We observe that at low pump pulse energies the resonance frequency shift increases linearly with pump-pulse energy, which agrees with the positive refractive index change of the electronic Kerr effect \cite{boyd.2008.book, yuce.2012.josab}. 
At high pump pulse energies, however, free carriers are excited that reduce the refractive index, opposite to the Kerr effect~\cite{yuce.2012.josab}. 
With our settings of pump and probe frequencies the free carriers can only be excited by two- and three-photon processes.
Since the probe pulse energy is small the non-degenerate two photon excitation of free carriers is negligible.
For this reason, the dependence of the refractive index (and hence resonance frequency) becomes nonlinear versus pump-pulse energy at high pump pulse energies. 
As shown in Fig. \ref{yuce-f6} the power of the pump pulse should be set such as to avoid free carrier generation via three-photon excitation. 
Beyond $70 \ pJ / \mu m^2$ pump pulse energy free carriers are excited, which effectively sets a limit to the pump pulse energy, and thereby to the frequency shift of the cavity resonance via the electronic Kerr effect.

\subsection{The effect of the quality factor} \label{SecQfactor}

\begin{figure}[htb]
  \centering
  \includegraphics[width=0.6\columnwidth]{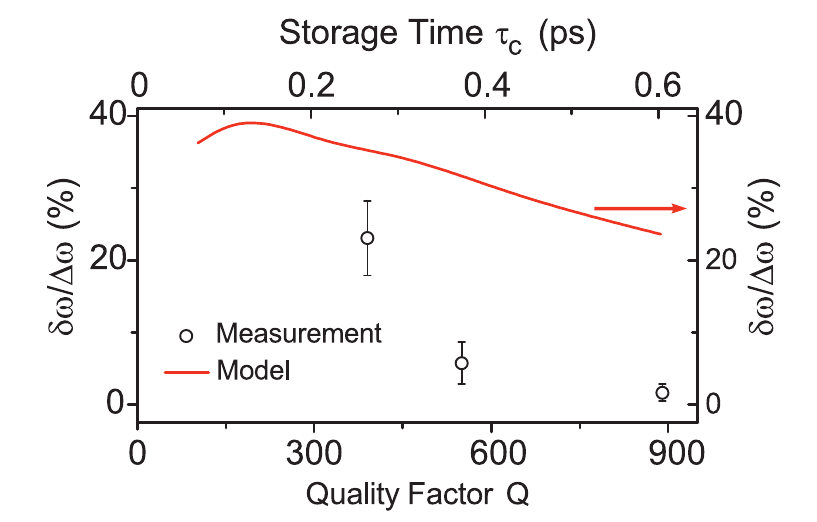}
  \caption{ Relative cavity resonance frequency change versus quality factor and cavity storage time. The calculations and the experiments are performed at $65 \ \rm{pJ/\mu m^2}$ pump fluence. Black circles show the measured results within the standard deviation. The solid curve indicates the calculated relative frequency change for different quality factor cavities.}
\label{yuce-f7}
\end{figure}

We have performed switching experiments on cavities with different quality factors to investigate the effect of the cavity storage time $\tau_c$ on the Kerr-induced resonance frequency change. 
Figure~\ref{yuce-f7} shows the relative cavity resonance frequency shift versus both the quality factor $Q$ and the storage time $\tau_c$ of the cavity. 
We observe that the shift of the cavity resonance frequency ($\delta \omega$) relative to the cavity linewidth ($\Delta \omega$) is maximal when the storage time is matched to the pump pulse duration $\tau_P$. 
We see both in our measurements and in our model that increasing the storage time $\tau_c$ of the cavity not only decreases the switching speed but also decreases the induced frequency shift induced via the Kerr effect. 
This can be understood since the magnitude of the observed frequency shift ($\delta\omega$) is given by the time-overlap integral of the pump and probe light that is stored in the cavity~\cite{ctistis.2011.apl}. 
The decreasing frequency shift with increasing quality factor is caused by the decreased temporal overlap of pump and probe as the cavity-stored probe pulse becomes much longer than the pump pulse ($\tau_{cav}\gg\tau_{P}$).  

In agreement with our experiments, our model in Fig.~\ref{yuce-f7} shows that a greater resonance frequency shift will be observed for a cavity that matches the switch pulse duration during the Kerr switching of a cavity. 
The relative shift of the resonance frequency is maximized at \fs{140} when the duration of the cavity-stored probe matches the pump duration ($\tau_c \simeq \tau_P$). 
Our model predicts a greater resonance frequency shift compared to our experiments. 
Our dynamic model in Fig.~\ref{yuce-f7} does not include the effect of the counteracting free carriers that is more pronounced for high quality factor cavities. 
In a cavity with a high quality factor the probe light intensity is enhanced and thereby the probability of degenerate and non-degenerate two photon excitation of free carriers is increased~\cite{yuce.2012.josab}. 
For this reason, the difference between our model and our experiments is larger for high quality factor cavities since we can not include free carrier effects in our dynamic model. 
The non-degenerate two photon excitation rates are calculated for steady state~\cite{yuce.2012.josab}. 
Cavities with a short storage time reduce the excited free carrier density and increase the temporal overlap of the ultra-short pump pulse with the cavity-stored probe light ($\tau_c \simeq \tau_P$), which even allows to repeatedly Kerr switch a cavity resonance at exhilarating THz clock rates~\cite{yuce.2013.ol}.

\begin{figure}[htb]
  \centering
  \includegraphics[width=0.55\columnwidth]{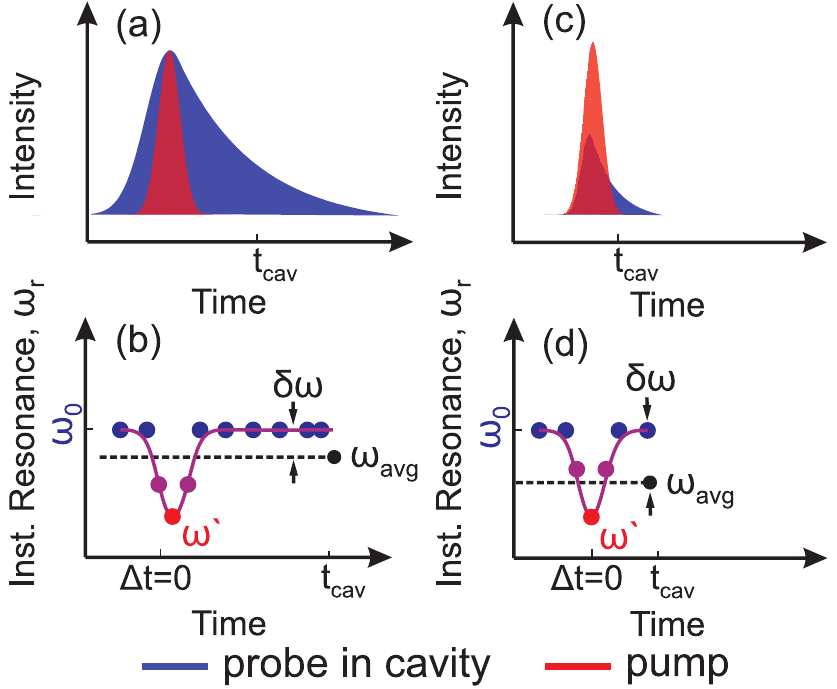}
  \caption{Schematic representation of the pump and probe pulses in the cavity for two different quality factor cavities. The lower panels show the instantaneous frequency shift versus time. The cavity resonance instantaneously shifts from $\omega_0$ to $\omega^\prime$ at pump probe overlap ($\Delta t=0$). The detected resonance shift ($\omega_{avg}$) is deduced from the transient reflectivity that is a result of the time averaging of the cavity storage and detector response time. A larger resonance frequency shift is observed for cavities with shorter storage times.}
\label{yuce-f8}
\end{figure}

Figure~\ref{yuce-f8} schematically illustrates the effect of the storage time of the cavity on the Kerr switching of a microcavity in real time. 
We plot situations where the delay $\Delta t$ is such that the overlap of the pump and cavity-stored probe is maximal, and we assume the observed cavity resonance $\omega_{avg}$ to be averaged over the whole pulse duration in view of the relatively slow detection ($t_{int} > \tau_{pr}$), see section~\ref{setup}.
For a high quality factor cavity, see Fig~\ref{yuce-f8}(a) and (b), there is no pump light during a long fraction of the probe pulse in the cavity since $\tau_c \gg \tau_P$. 
As a result, the average resonance frequency shift $\omega_{avg}$ is small. 
Given the same refractive index change, one would naively expect to observe a greater relative shift of the cavity resonance for a high quality factor cavity. 
However, this is only true if the switch duration is longer than the cavity storage time ($\tau_c < \tau_P$). 
Since the Kerr switching of the cavity is performed within the pump pulse duration, one has to consider the overlap integral in time to get the average resonance frequency shift. 
This overlap is determined both by the duration of the pump pulse and the quality factor of the cavity that affects the duration of the cavity-stored probe pulse. 
Consequently, at similar switch conditions a cavity with a shorter storage time will reveal a greater shift of the time averaged resonance $\omega_{avg}$ since the instantaneous cavity resonance shift $\omega^\prime$ has a larger weight. 
As illustrated in Fig.~\ref{yuce-f8} (c) and (d), a larger portion of the cavity-stored probe light overlaps with the short pump pulse in a cavity that has a short storage time, close to the pump pulse duration ($\tau_c \simeq \tau_P$).

\subsection{The effect of the pump pulse duration}

\begin{figure}[htb]
  \centering
  \includegraphics[width=0.6\columnwidth]{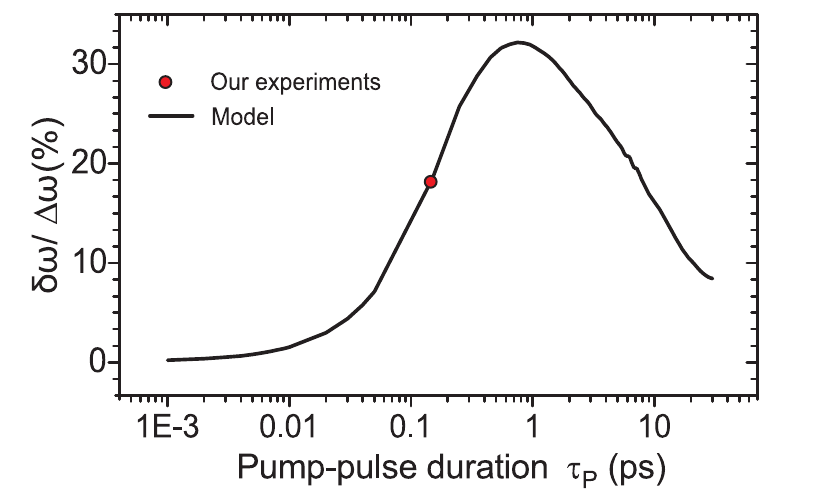}
  \caption{ Relative cavity resonance frequency change with respect to the cavity linewidth ($\Delta \omega$) versus the pump pulse duration. The red circle marks the duration of the pump pulse in our experiments.}
\label{yuce-f9}
\end{figure}

To investigate the effect of the pump pulse duration on Kerr-induced cavity resonance frequency switching, we have performed calculations on a switched cavity at different pump pulse durations using our dynamic model. 
Similar to our experiments, the microcavity in our calculations has 7 pairs of GaAs/AlAs layers in the top mirror and 19 pairs of GaAs/AlAs layers in the bottom mirror and is surrounded by air. 
For the nanostructure described we get a quality factor of $Q=450$ in our calculations whereas we measure a quality factor of $Q=390$. 
The difference in measured and calculated quality factors are due to absence of loss mechanisms, such as slight deviations of the layer thicknesses versus the nominal design.
Moreover, in our model we do not include GaAs wafer, which also increases the quality factor, given the increased contrast between air and the bottom Bragg mirror.
The calculations are performed at a pump fluence of $65 \ \rm{pJ/\mu m^2}$ and the peak intensity is kept constant for each pulse duration. 
Figure~\ref{yuce-f9} shows the cavity resonance frequency shift relative to the resonance linewidth ($\delta \omega / \Delta \omega$) versus the pump pulse duration at pump probe delay $\Delta t=0$. 
The maximum shift ($\delta \omega$) reaches $32\%$ when the pump pulse duration is set to $\tau_P = \fs{550}$ for this particular cavity.
After $\tau_P= \ps{1.1}$ the cavity resonance frequency shift starts to decrease with increasing pump pulse duration.
The optimum pump pulse duration is in the range of the cavity storage time ($\tau_{c}=\fs{300}$).

\begin{figure}[htb]
  \centering
  \includegraphics[width=0.9\columnwidth]{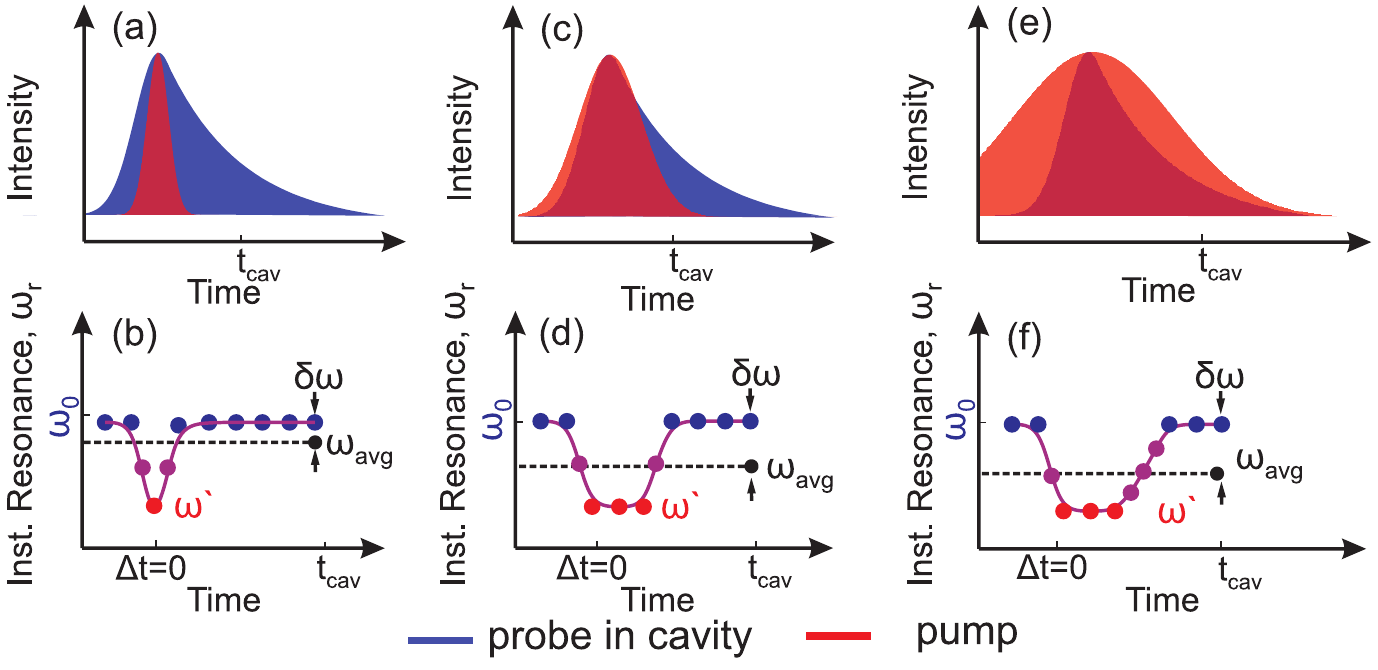}
  \caption{Schematic representation of the pump and probe pulses in the cavity for three different pump pulse durations. The peak intensity of the pump pulse is kept constant while stretching the pump pulse. The lower panels show the instantaneous frequency shift versus time. The cavity resonance instantaneously shifts from $\omega_0$ to $\omega^\prime$ at pump probe overlap ($\Delta t=0$). The detected resonance shift ($\omega_{avg}$) is deduced from the transient reflectivity that is a result of the time averaging of the cavity storage and detector response time. A larger resonance frequency shift is observed if the pump pulse duration is matched to the cavity storage time.}
\label{yuce-f10}
\end{figure}

Figure~\ref{yuce-f10} schematically depicts the probe pulse that is in resonance with the cavity and the pump pulse. 
The resonance frequency of the cavity shifts from $\omega_0$ to $\omega^\prime$ due to the instantaneous change of the refractive index. 
Given the time averaging of the detector, we observe an average resonance frequency shift $\omega_{avg}$ that is smaller than the instantaneous shift.
For a short pump pulse duration ($\tau_P < \tau_c $) as illustrated in Fig.~\ref{yuce-f10}(a) the cavity-averaged frequency shift is small, see Fig.~\ref{yuce-f10}(b). 
As the pump pulse gets longer in time (Fig.~\ref{yuce-f10}(c)) the weight of instantaneous shift increases in the time averaged resonance frequency change. 
As a result, a greater shift of the cavity resonance is observed, see Fig.~\ref{yuce-f10}(d). 
Stretching the pump pulse much longer than the cavity storage time tomographically samples the probe light in the cavity, see Fig.~\ref{yuce-f10}(e). 
The average resonance frequency shift decreases when $\tau_P \gg \tau_c $ since the magnitude of the frequency shift $\delta\omega$ is given by the overlap integral of the pump and probe~\cite{ctistis.2011.apl}. 

Our calculations show that using the same cavity the relative resonance frequency shift can be increased from $\delta\omega / \Delta\omega=19\%$ to a maximum of $\delta\omega / \Delta\omega=32\%$ only by increasing the pump pulse duration from $\tau_{pu}=\fs{140}$ to $\tau_{pu}=\fs{550}$. 
Furthermore, using a AlGaAs cavity the shift can be increased three times to $\delta\omega / \Delta\omega \simeq 100\%$ (or one linewidth) given its large nonlinear refractive index coefficient~\cite{caspani.2011.josab}. 
In this case, for a cavity operating at telecom wavelengths, the pump frequency should be set to about $\icm{7282}$ to maximize the third order susceptibility via non-degenerate pump and probe photons. 
Two photon excitation of free carriers via degenerate pump photons will still be suppressed at this photon energy. 
Here, we note that cavities with short storage times are required for fast switching times and also for avoiding probe intensity enhancement, which can lead to unwanted free carrier excitation via two probe photons.


\section{Conclusion}

We have studied the ultrafast all-optical switching of GaAs/AlAs and AlGaAs/AlAs semiconductor microcavities at telecom wavelengths using the electronic Kerr effect. 
We show that the judicious tuning of the amplitude and frequency of the driving fields relative to the bandgap of the semiconductor decreases the number of free carriers and also increases the positive shift of the resonance frequency resulting from the electronic Kerr effect.
We investigate the effect of the temporal overlap of the pump and the probe in Kerr switching experiments as a function of pump pulse duration and quality factor. 
High-Q cavities invite Kerr switching with much longer pump pulses. 
We show that the refractive index change induced by the electronic Kerr effect is increased at the maximum temporal overlap of the pump and the probe pulses. 
The realization and the understanding of the time evolution and the dependency of the Kerr effect to material parameters reveal the set of parameters using which the shift of a cavity resonance can be increased to one linewidth using the electronic Kerr effect. 

\section{Acknowledgements}
We thank Allard Mosk and Henri Thyrrestrup for useful discussions. 
This research was supported by FOM-NWO, the NWO-Nano program, ERC-Pharos, and STW. 
\end{document}